\pdfoutput=1
\documentclass[journal=langd5,manuscript=article]{achemso}

\usepackage{graphicx}
\usepackage{dcolumn}
\usepackage{bm}
\usepackage[version=3]{mhchem}
\usepackage[per-mode=symbol]{siunitx}
\usepackage{booktabs}

\DeclareSIUnit\rydberg{Ry}
\DeclareSIUnit\atmosphere{atm}
\DeclareSIUnit\calorie{cal}

\author{Carlos Alberto Martins Junior}
 \affiliation[Universidade de Sao Paulo]{Universidade de Sao Paulo, Instituto de Fisica, Rua do Matao 1371, Sao Paulo, SP, 05508-090, Brazil}
 \email{camjjr.cm@usp.br}
\author{Henrique Musseli Cezar}
 \affiliation[Universidade de Sao Paulo]{Universidade de Sao Paulo, Instituto de Fisica, Rua do Matao 1371, Sao Paulo, SP, 05508-090, Brazil}
 \alsoaffiliation[Hylleraas Centre]{Hylleraas Centre for Quantum Molecular Sciences and Department of Chemistry, University of Oslo, PO Box 1033 Blindern, 0315 Oslo, Norway}
 \email{h.m.cezar@kjemi.uio.no}
\author{Daniela Andrade Damasceno}
 \affiliation[Universidade de Sao Paulo]{Universidade de Sao Paulo, Instituto de Fisica, Rua do Matao 1371, Sao Paulo, SP, 05508-090, Brazil}
 \email{daniela.damasceno@usp.br}
\author{Caetano Rodrigues Miranda}
 \affiliation[Universidade de Sao Paulo]{Universidade de Sao Paulo, Instituto de Fisica, Rua do Matao 1371, Sao Paulo, SP, 05508-090, Brazil}
 \email{crmiranda@usp.br}

\title{Effects of van der Waals interaction on the \ce{N2} adsorption on carbon nanotubes: proposal of new force field parameters}
\keywords{Molecular Dynamics,Carbon Nanomaterials, Graphene, \ce{N2},\ce{CO2}, Monte Carlo, Monte Carlo Simulation, potential}

\begin{document}


\section{Abstract}
The separation of carbon dioxide (\ce{CO2}) from nitrogen gas (\ce{N2}), the main component of flue gas, has become an emerging action to mitigate climate change. Feasible and efficient approaches to exploring the separation properties of materials are molecular dynamics (MD) and Monte Carlo (MC) simulations. In these approaches, a careful choice of force fields is required to avoid unrealistic predictions of thermodynamic properties. However, most studies use Lorentz-Berthelot combining rules (LB) to obtain the interaction between different species, an approximation that could not capture the essence of interfacial interactions. In this context, we verified how accurate LB is in describing the interaction of \ce{N2} molecules and carbon nanostructures by comparing the interaction energies from LB with those from density functional theory (DFT) calculations. We selected carbon nanomaterials because they are considered promising materials to perform \ce{N2}/\ce{CO2} separation. The results show that the LB underestimates the interaction energies and affects the prediction of fundamental properties of solid-fluid interfacial interactions. To overcome this limitation, we parametrized a Lennard-Jones potential using energies and forces from DFT, obtained through the van der Waals functional KBM. The proposed potential show good transferability and agreement to \textit{ab-initio} calculations. Grand Canonical Monte Carlo simulations were performed to verify the effects of employing LB in predicting the amount of nitrogen gas adsorbed inside different CNTs. LB predicts a lower density inside them. Moreover, our results suggest that LB leads to a different characterization of the adsorption properties of carbon nanotubes, by changing significantly the adsorption isotherm.


\section{Introduction}
One of the most studied approaches to mitigate carbon dioxide emission (\ce{CO2}) is Carbon Capture and Storage (CCS)\cite{gibbins2008}, which consists of capturing greenhouse gases from the combustion of fossil fuels and storing it in geological formations. However, to perform CCS, it is necessary to efficiently separate \ce{CO2} from other molecules that compose flue gases. These gases are mainly composed of nitrogen gas (\ce{N2}) \cite{granite2002,Eon2003}, which has driven several studies to improve the separation process of this gas from carbon dioxide \cite{jia2007}.

A class of materials that possesses desirable properties for gas separation application is carbon nanostructures\cite{sanip2011,sanip2011-2}. They are mechanically and chemically stable \cite{yan2012_2,Galashev_2014} and, besides that, their properties can be easily tuned by functionalization, i.e., by incorporating some radicals in the structures \cite{sanip2011-2}, or by doping it \cite{lee2018_2}. For example, studies have found that functionalized multi-walled carbon nanotubes enhance the separation properties of mixed matrix membranes \cite{LEE2018159}. In another study, experiments showed that nanoporous graphene (NPG) separates different gases \cite{koenig2012}, which boosted the investigation of the separation properties of NPGs. For instance, in the case of \ce{N2}/\ce{CO2} separation, nanoporous with different diameters and geometries were tested \cite{wang2016,wang2019}. Moreover, different atoms were used to passivate the nanoporous rim \cite{wang2016,liu2013}. These studies have shown that NPGs can efficiently perform the \ce{N2}/\ce{CO2}  separation, reaching a selectivity of 100\%.  

Feasible and efficient approaches to study gas separation are Molecular Dynamics or Monte Carlo simulations. In these types of molecular modeling, the interactions between different species are usually obtained by a Lennard-Jones potential, 
\begin{equation}
    V(r)=4\epsilon \left[ \left(\frac{\sigma}{r}\right)^{12}-\left(\frac{\sigma}{r}\right)^{6}\right]
\end{equation}
in which the $\epsilon$ and $\sigma$ parameters are given by averaging the parameters that describe the isolated species. A common way of performing such averages is through the Lorentz-Berthelot combination rules (LB), $\sigma_{ij}=(\sigma_{ii}+\sigma_{jj)}/2$ and $\epsilon_{ij}=\sqrt{\epsilon_{ii}\epsilon_{jj}}$, in both of which the sub-index \textit{ij} refers to the cross interaction while \textit{ii} and \textit{jj} refer to the pure species. Although LB are widely employed, different studies have shown that it may not capture the essence of interfacial interactions \cite{co2-trabalho,inadequacy_lb}.

In the case of \ce{N2} interacting with graphene, there is at least one force field that goes beyond LB: Vekeman {\it et al}. proposed new parameters derived from density functional theory (DFT) calculations \cite{vekeman2018-2}. However, the chosen functional, B97-D, is known to significantly deviate from coupled clusters calculations (CCSD(T)), which is considered to fully describe van der Waals interactions. For example, a deviation of \SI{18.25}{\percent} was found for coronene adsorbed on graphene \cite{yeamin2014}. Recent studies confirm the accuracy of van der Waals functionals, such as KBM and C09 \cite{KBM,C09}, to describe adsorption energies. For instance, results show that the adsorption energy of \ce{CO2} on graphene obtained from both functionals agree with experimental results \cite{co2-trabalho,Takeuchi2017}. In particular for \ce{N2} and carbon nanomaterials, such functionals should be employed since vdW forces are the main interaction \cite{shojaie2018}. Besides the functional, another consideration taken in the study performed by Vekeman {\it et al.} that might affect the final results is the use of a finite size material to perform the calculations instead of using a graphene sheet. This was probably done due to a limitation of the used software. However, the finite size and strategies used to passivate the border of the material that mimics graphene may affect the adsorption energies.

Based on the above review of the literature, we compared the adsorption energy curves of \ce{N2} molecules on carbon structures obtained from different classical force fields (FFs) using LB rules with the ones from DFT calculations using the KBM functional. A large discrepancy between results from LB and DFT was observed, which imposed a need for developing a precise potential with a new set of cross parameters for the \ce{N}\textendash \ce{C} and \ce{N}\textendash \ce{H} interactions. The proposed potential showed high accuracy in describing the adsorption energy of \ce{N2} molecules on graphene and carbon nanotubes compared to DFT. To verify the effects of the underestimation of the vdW interactions on the thermodynamics properties, we performed Monte Carlo simulations to obtain the load of \ce{N2} inside carbon nanotubes, at \SI{300}{\kelvin} and \SI{1}{\atmosphere}, and the adsorption isotherm at \SI{300}{\kelvin}. The results reveal that LB underestimates the load, independent of force field used to obtain the interaction, and change the adsorption properties, since the isotherms obtained from the proposed potential are significantly different to the one obtained through LB. These findings indicate that LB might not be adequate for the full description of the interaction of different species when vdW forces play an important role, as in the case of \ce{N2} and carbon nanomaterials, and lead to results inconsistent with those obtained with electronic dispersion and experiments present in the literature.

\section{Methods and Models}

\subsection{Carbon Nanostructures}
We considered the following carbon nanostructures: a $5\times5$ graphene sheet and $(m,n) = (8,8), (10,10), (0,17)$ carbon nanotubes. These structures were created using the VMD software \cite{vmd} and the Topotools plug-in \cite{topotools}. It was decided not to optimize the other nanostructures to match the bond lengths with those from the classical force fields. For the \ce{N2} bond length, we used a value of \SI{1.10}{\angstrom}, while for the carbon nanomaterials, bond lengths were considered to be \SI{1.418}{\angstrom}.

\begin{figure}
\includegraphics[width=0.7\columnwidth]{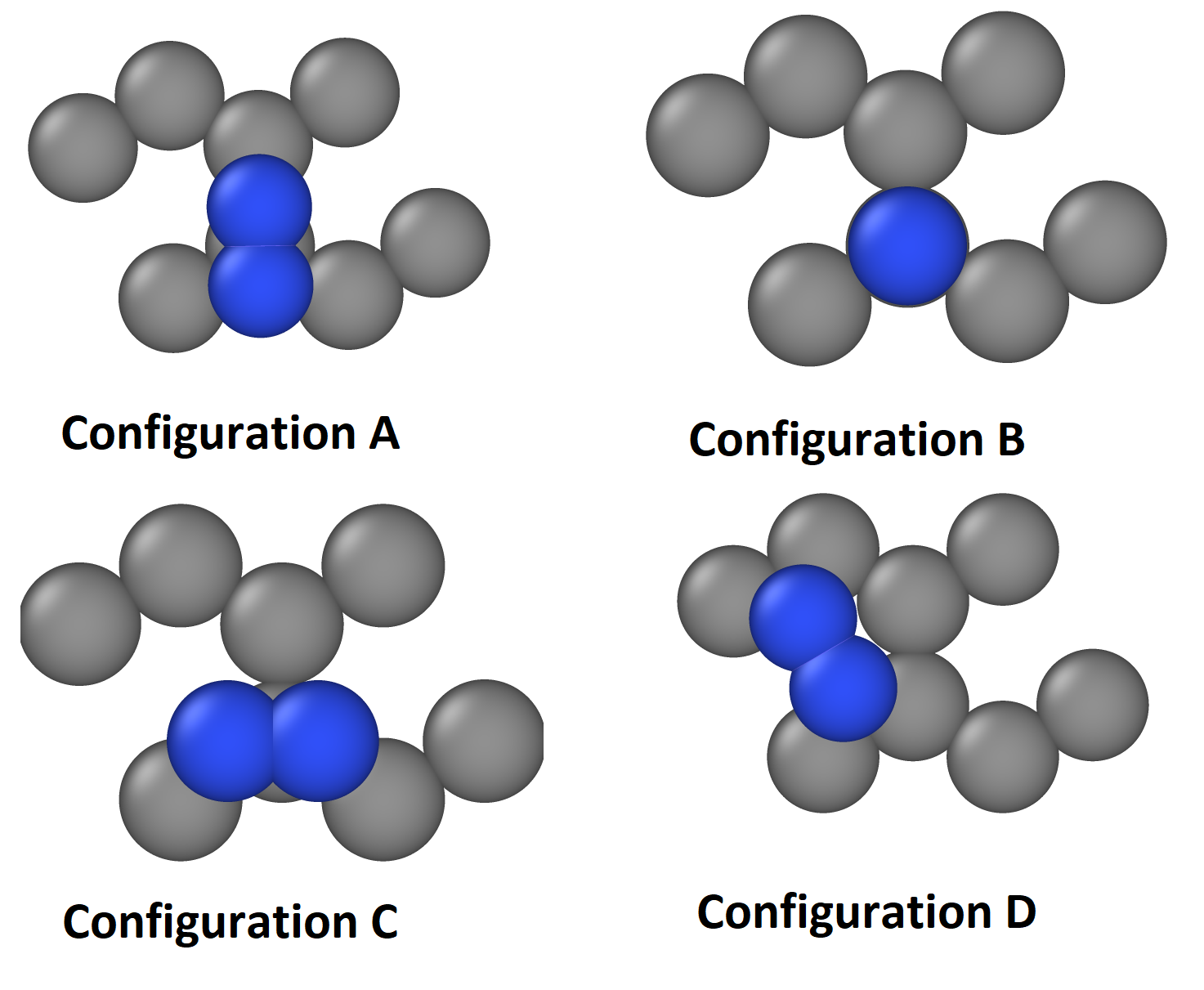}
\caption{\label{fig:configs} Adsorption configurations considered in the calculations. For clarity, these images show a nitrogen molecule adsorbed on a $2\times2$ graphene cell. The gray spheres represent the carbon atoms, the blue ones represent the nitrogen atoms, and the white ones represent the hydrogens. The visualizations were generated using Ovito\cite{ovito}.}
\end{figure}

\subsection{Classical Force Fields}
To determine the accuracy of LB combination rules in describing the interaction between \ce{N2} and carbon nanostructures, we compared the interaction energies obtained using different classical FFs with those from DFT calculations. We used the AIREBO \cite{airebo}, Amber \cite{amber}, Mao \cite{Mao_1999}, Huang \cite{huang}, Steele \cite{steele}, and Walther \cite{walther} force fields for the carbon structures and the Trappe model for \ce{N2}. They all use the Lennard-Jones potential, with $\epsilon_{ii}$ and $\sigma_{ii}$ presented in Table \ref{tab:force_fields}. We selected only a model for \ce{N2} due to the small differences in the force field parameters of different models, as can also be seen in table \ref{tab:force_fields}. The interaction energies were obtained in four different adsorption sites. The adsorption configurations are represented in figure \ref{fig:configs}. 
\begin{table}[ht]
    \centering
    \caption{Force field parameters for carbon structures (first part of the table) and nitrogen molecules (second part of the table). All force fields for \ce{N2} studied here are three-site models. In such models, two sites are localized in each nitrogen atom while the other is localized at the center of the molecule. The charge $q_i$ is the value at the nitrogen atoms. For the site at the center of the molecule, the charge is $-2q_i$. \label{tab:force_fields}}
    \begin{tabular}{@{}cccc@{}}
 \toprule
 Force Field     &  $\epsilon (K) $ & $\sigma $ (\AA)  & $q_i (e) $ \\
 \midrule
 AIREBO \cite{airebo}     &  $32.96$ & $3.400$  &  -\\
 Amber \cite{amber}     &  $43.30$ & $3.400$  &  -\\
 Mao \cite{Mao_1999}     &  $48.78$ & $3.370$  &  -\\
 Huang \cite{huang}    &  $35.26$ & $3.550$  & -\\
 Steele \cite{steele}    &  $28.00$ & $3.400$  & - \\
 Walther \cite{walther}     &  $52.87$ & $3.851$  & - \\
 \midrule
 Trappe \cite{trappe} & 36.00 & 3.31& -0.482\\
 Garcia-Perez \cite{garcia-perez2017} & 36.40 & 3.32& -0.40484\\
 Vujíc \cite{vujik} & 40.24 & 3.32& -0.482\\
 Murthy \cite{murthy}& 36.44 & 3.33& -0.482\\
 \bottomrule
\end{tabular}
\end{table}

\subsection{DFT calculations}
\label{subsection:DFT}

DFT calculations were performed using the KBM exchange-correlation functional \cite{KBM}. This functional was found to successfully describe different systems in which the main interaction is due to van der Waals forces \cite{KBM,hensley2017}, as for \ce{N2} molecules interacting with carbon materials\cite{shojaie2018}. Furthermore, recent studies have shown that the KBM functional gives good results for the adsorption energy of \ce{CO2} molecules on graphene, a system similar to the one investigated here \cite{Takeuchi2017,co2-trabalho}. We also tested other vdW and non-vdW functionals, namely, the C09 \cite{C09} and BH \cite{BH}, an LDA functional, PW92 \cite{PW92}, and a GGA functional, PBEsol \cite{PBEsol}. These tests did not show a significant difference between the vdW functionals, with the largest difference in the depth of the well being from KBM to C09, \SI{0.4424}{\kilo\calorie\per\mol} (\SI{7.2}{\percent}). However, some significant differences were found from the vdW functionals to PW92 and PBEsol (see Figure~S1 in the Supporting Information). Even though the LDA functional can describe the depth of the interaction energy, when compared to the vdW functionals, it cannot describe all the features of the interaction energies curves. These tests also show that GGA - PBEsol cannot describe either the depth of the interaction curve or the other parts. This pattern indicates the importance of including electron dispersion when studying such systems. 

We used a plane-wave cutoff of \SI{400}{\rydberg} for the grid and a convergence criterion of \num{e-4} in the elements of the density matrix. The selected basis was a double zeta plus polarization orbitals, with the second zeta carrying \num{0.35} of the norm. Based on convergence tests, we selected an $8\times8\times1$ k-points grid for the single-point calculations involving the $5\times5$ graphene sheet and $1\times1\times1$ k-point for the CNTs. All the DFT calculations considered periodic boundaries conditions. Thus, to avoid interaction between the images of the system, the distance between the graphene sheets was \SI{50}{\angstrom} and the CNTs were placed in the center of a \SI{20}{\angstrom} $\times$ \SI{20}{\angstrom} plane. We also tested the convergence of the interaction energy on the length of the carbon nanotubes, which is present in Figure~S2 of the Supporting Information. Based on these calculations, a length of \SI{7.368}{\angstrom} was selected for the arm-chair nanotubes and \SI{8.508}{\angstrom} for the zig-zag nanotube. All the dft calculations were performed using the SIESTA code REF.

The interaction energies from DFT were obtained using Eq. \ref{eq:interaction_energy}. The term $E_\text{int}$ is the interaction energy, $E_\text{sys}$ is the energy of the whole \ce{N_2}/carbon-nanomaterial system, $E_\text{c}$ and $E_{\ce{N2}}$ are the energies of the isolated carbon nanomaterial and isolated \ce{N2} molecule, respectively.

\begin{equation}
E_\text{int}=E_\text{sys}-E_\text{c}-E_{\ce{N2}}
\label{eq:interaction_energy}
\end{equation}

\subsection{Classical Molecular Simulations}
The interaction energies for the classical force fields were obtained using LAMMPS \cite{LAMMPS}, and the  Moltemplate software \cite{Moltemplate} was used to create the inputs. The energies were calculated using the compute group/group command, which does not include the interactions between the periodic images. Consequently, there was no need to test the convergence of the adsorption energy on CNT length, as we did for the DFT calculations. Also, since we used classical force fields that use Lennard-Jones potential, which is a pairwise potential, an effect due to many-body interactions between the periodic images does not exist, as might happen in the DFT calculations. 

Grand canonical Monte Carlo (GCMC) simulations were performed using the Cassandra software \cite{cassandra} in order to obtain the \ce{N2} density inside different carbon nanotubes using the force fields tested here and the fitted potential. To obtain the chemical potential of \ce{N2} at \SI{300.0}{\kelvin} and \SI{1.0}{\atmosphere}, we used the Widom insertion method\cite{widom}. These simulations were carried for \num{4e6} MC steps, with the last \num{3e6} steps being considered for the production phase. Then, using the obtained chemical potential we performed GCMC simulations to load \SI{200}{\angstrom} long CNTs with \ce{N2}. Four different CNTS were considered: (6,6), (8,8), (10,10) and (12,12). Initially, \num{5e4} simulation steps in the NVT ensemble were performed to obtain the initial configuration. Then, a GCMC simulation of \num{5e6} MC steps was performed, in which molecules were translated, rotated, inserted, or deleted inside the carbon nanotubes with a \SI{25}{\percent} probability for each move. For this last simulation, we employed configurational-biased insertions with \num{16} trial insertions. A cutoff of \SI{15}{\angstrom} was used for the Lennard-Jones plus Coulomb potential along with the Ewald method for long-range electrostatics, in which accuracy of \num{e-5} was selected.  

To verify the effects of the different force fields on adsorption properties, we also used GCMC simulations to obtain adsorption isotherms of CNT (10,10) and (0,17) at $T=\SI{300}{\kelvin}$, using the Steele's, Walther's and the fitted force field. We selected the Steele force field since it predicts the smallest density on CNT (10,10), as can be seen in the Section Results. For these simulations, we first simulated bulk \ce{N2} with different chemical potentials at the GCMC ensemble to obtain the average value of pressure. Simulations of \num{3.5e6} MC steps were performed to reach the equilibrium and, then, \num{3e6} MC steps were performed for the production phase. We used a tail correction for the simulations of the bulk, while the other simulations parameters were the same as previously discussed. Then, other simulations were performed to insert \ce{N2} molecules inside the carbon nanotubes. With these last simulations, we matched the number of adsorbed molecules inside the CNTs to the bulk pressure, obtaining the adsorption isotherm. Aside from the tail correction, we used the same simulation parameters. For the calculations of the gas density, we used the accessible volume, which was found by monitoring the position of nitrogen atoms inside the CNTs.           

\subsection{Fitting Procedure}
Due to the difference between the energy curves from the DFT calculations and LB rules (see the Results section), we fitted a new set of Lennard-Jones parameters for the interactions between nitrogen and (10,10) and (0,17) CNTs interfacial interactions using GULP \cite{gulp}. To better fit the minimum of the energy curves, a weight based on the Boltzmann distribution was used,
\begin{equation}
    w=\mathrm{e}^{\frac{E-E_\text{min}}{kT}}
    \label{eq:}
\end{equation}
in which $w$ is the weight, $E_\text{min}$ is the minimum of the curve, $k$ is the Boltzmann constant, and $T=\SI{350}{\kelvin}$ is the temperature. It is worth mentioning that $T$ does not hold a physical meaning and is just used to obtain a better description around the curve minimum. We also added another weight to the energies to balance the fact that the number of forces is greater since there are $3N$ forces components, being $N$ the number of atoms, for each point in the interaction energy curve. The temperature and energy weight were selected by manually checking how well the fitting was compared to DFT energies and forces. 

\section{Results and Discussion}
\subsection{Comparison between DFT and FFs}
\begin{figure}[ht]
\includegraphics[width=\columnwidth]{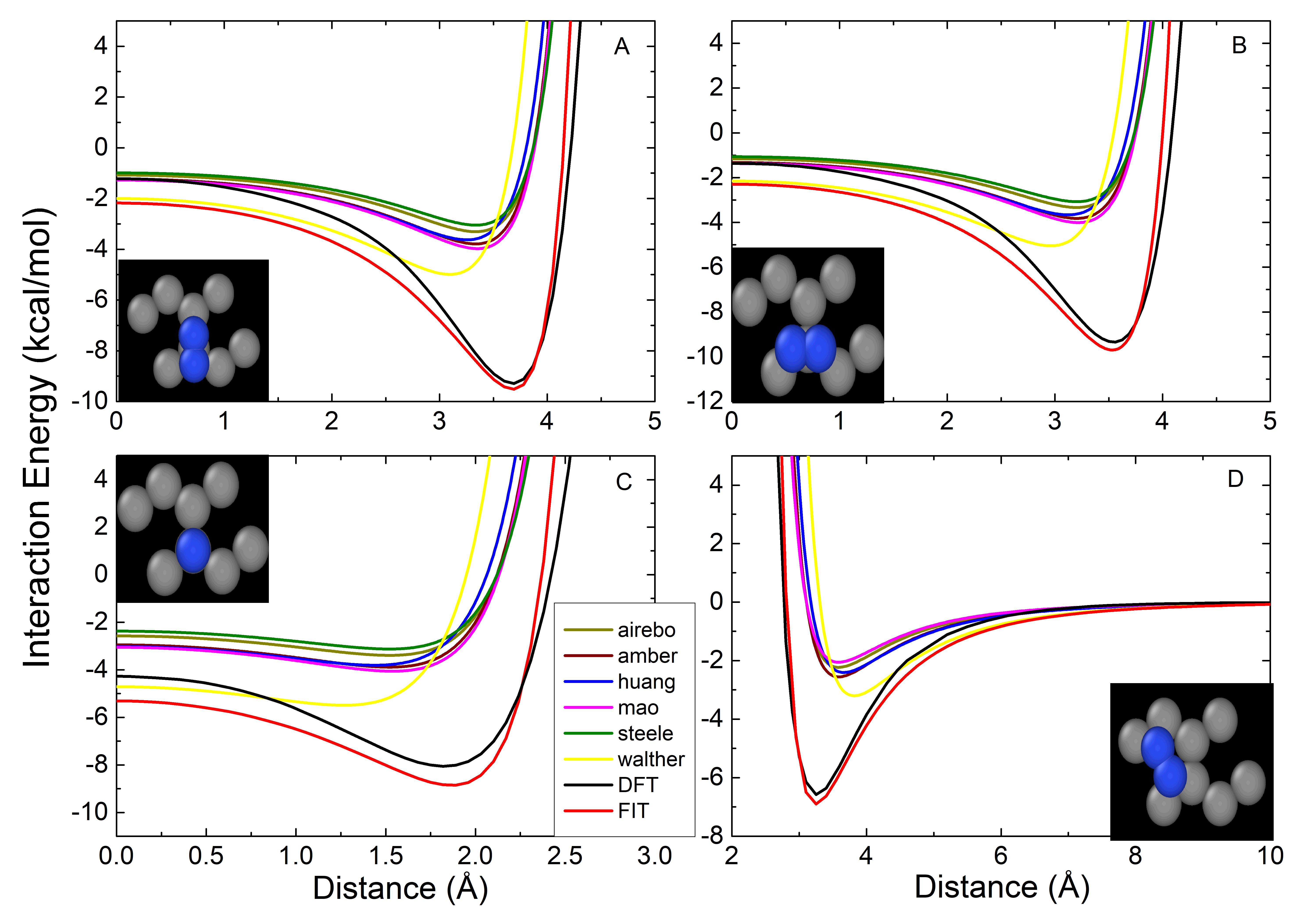}
\caption{\label{fig:dft_versus_fit} A) (10,10) CNT, B) (0,17) CNT, C) (8,8) CNT, and D) graphene. Comparison of the interaction energies of \ce{N2} and different carbon nanostructures, using DFT-KBM, fitted LJ parameters, and classical forces fields through Lorentz-Berthelot combination rules. The fitting was performed using data from the (10,10) and (0,17) CNTs. The curves for graphene and the (8,8) CNT show the transferability of the potential to both smaller and larger curvatures.}
\end{figure}

Figure \ref{fig:dft_versus_fit} shows the adsorption energies curves obtained with DFT, the classical force fields using LB combination rules, and the proposed potential. These results indicate that the classical force fields underestimate the interaction energy between the nitrogen molecule and graphene or carbon nanotubes. Neither the position nor the depth of the potential wells are properly described. Since the parameters for both isolated systems can describe their interactions, we attribute this discrepancy to LB combination rules. These rules are an approximation, and there is no guarantee they can be used to successfully describe interfacial interactions. 

The results also indicate that the adsorption site B (see Figure \ref{fig:configs}) is energetically the least favorable configuration and, aside from it, only slight differences between the other adsorption curves were observed, indicating no clear preferable adsorption site. A similar result was reported in the literature. Using molecular dynamics simulations, Vekeman {\it et al.} has recently shown that if the flexibility of the graphene sheet is taken into account, a united-atom model for nitrogen molecules gives almost the same results as an atomistic model \cite{vekeman2019-2}. This suggests that the effect of orientation of nitrogen molecules is negligible. Moreover, it is possible to conclude that vdW interaction does not play a role in the orientation since the results from KBM and their results, derived from the B97-D functional \cite{vekeman2018-2}, which does not include electron dispersion, give the same result for non-preferable orientation of \ce{N2} molecules on graphene. 

Some significant differences in the adsorption energy from our DFT calculations and others reported in the literature were found. Here, it was found an adsorption energy of \SI{-6.6}{\kilo\calorie\per\mol} of \ce{N2} on graphene, while a value of \SI{-2.6}{\kilo\calorie\per\mol} was reported using the B97-D functional \cite{vekeman2018-2} and a value of \SI{-3.2}{\kilo\calorie\per\mol} was reported using $\omega$B97X-D/cc-pVDZ \cite{gordeev2013}. We attribute these differences to the different levels of calculation employed in each work. First, in the cited studies, periodic boundary conditions were not used, which might affect the adsorption energies, since the finite size and strategies used to passivate the graphene border may affect the results. Second, it is known that the methodology employed in such studies leads to energies that are different from the ones obtained with coupled clusters calculations (CCSD(T)), which fully describes van der Waals interactions \cite{klimes2012}. For example, a deviation of \SI{18.25}{\percent} between DFT - B97-D and CCSD(T) was observed for coronene adsorbed on graphene \cite{yeamin2014}. Furthermore, the KBM exchange-correlation functional employed here was found to successfully describe the adsorption energy of carbon dioxide on graphene\cite{Takeuchi2017,co2-trabalho} 

Unfortunately, based on the authors' knowledge, there are no experimental data for \ce{N2} adsorption energies on graphene, as in the case of \ce{CO2}. The only experimental results available are for the adsorption energy of \ce{N2} on graphite, in which values of \num{-2.236} to \SI{-2.624}{\kilo\calorie\per\mol} have been reported for the depth of the well \cite{experimental_results_graphite}. However, the adsorption energy on graphite can be significantly different from graphene \cite{Takeuchi2017,co2-trabalho}.

\subsection{Fitting}

Due to the reported underestimation of the adsorption energies, we performed a fit of the Lennard-Jones cross parameters. As described in the methodology, we used only forces and energies from the (10,10) and (0,17) CNTs in the fitting process. The fitted LJ cross parameters are shown in Table \ref{tab:parameters}. Figure \ref{fig:dft_versus_fit} shows curves for the interaction energies obtained from DFT, the fitted potential, and classical forces fields employing the Lorentz-Berthelot combination rules. The results show that the differences between the proposed fit and the DFT calculations are smaller than \SI{1.0}{\kilo\calorie\per\mol}, which is within the chemical accuracy. Extending the application to other models, including graphene and (8,8) CNT, the results indicate that the fit is transferable for both smaller and larger curvatures.

\begin{table}[ht]
    \centering
    \caption{Fitted Lennard-Jones parameters for \ce{N-C} interactions.}
    \begin{tabular}{@{}ccc@{}}
    \toprule
         & $\epsilon_{ij}$ (K) & $\sigma_{ij}$ (\AA) \\
    \midrule     
    \ce{N-C}     & 118.82 & 3.011 \\
    \bottomrule
    \end{tabular}
    \label{tab:parameters}
\end{table}

\begin{figure}[ht]
\includegraphics[width=\columnwidth]{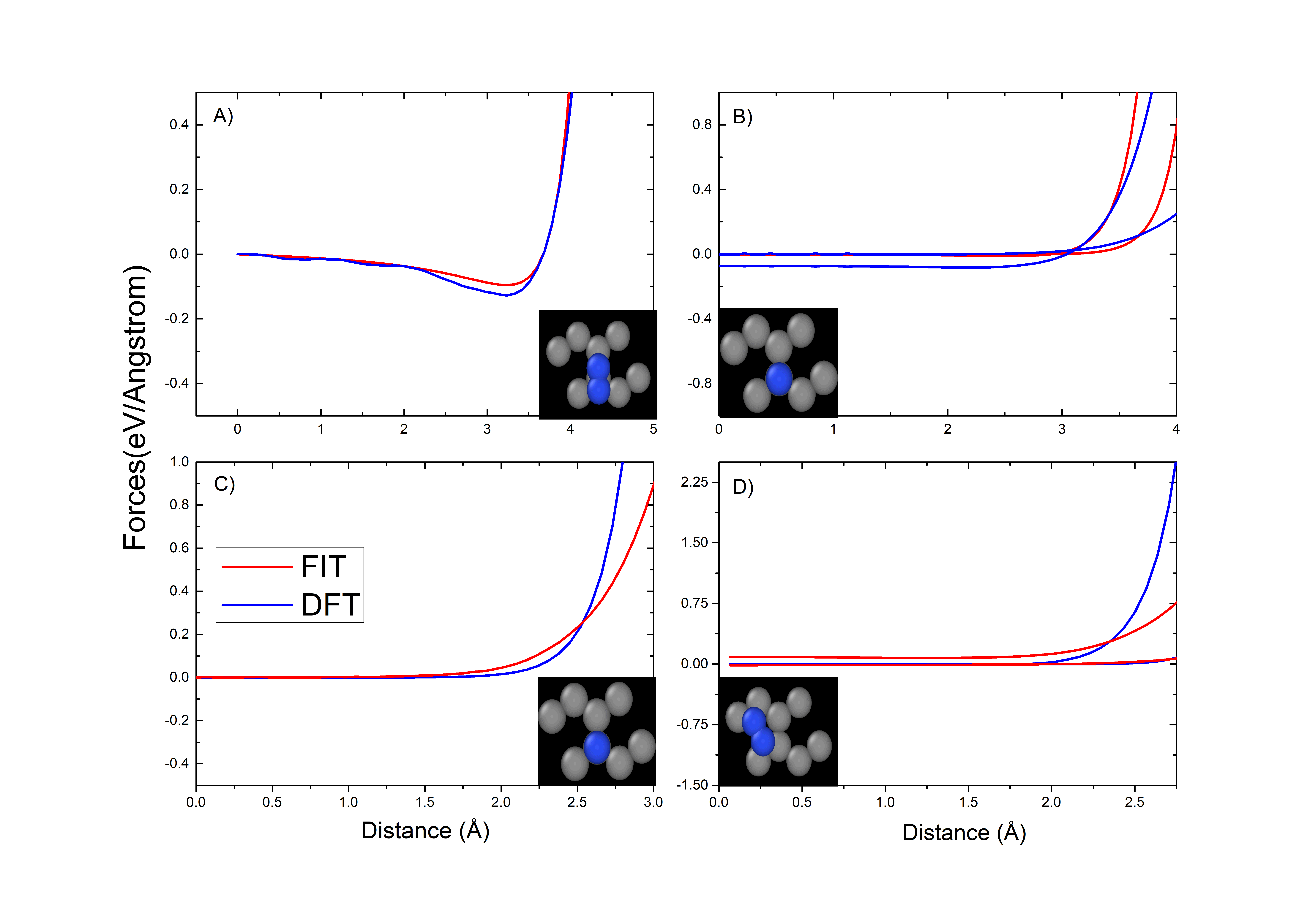}
\caption{\label{fig:forces_dft_versus_fit} Figures A and C show the component of the force on a \ce{N} atom in the direction that \ce{N2} is moving towards the carbon atoms. Figures B and D show the components of the force on a \ce{C} atom in the other directions. Figures A and B are from CNT(10,10) while figures C and D are from CNT(8,8). All figures show the comparison of the forces between \ce{N2} and carbon nanostructure, using DFT-KBM and our fitted LJ parameters.  }
\end{figure}

Figure \ref{fig:forces_dft_versus_fit} compares forces acting on selected atoms using fitted parameters and DFT calculations. Overall, the curves are in reasonable agreement. The major discrepancies in forces are when the nitrogen molecule is near the carbon atoms, as shown in Figures \ref{fig:forces_dft_versus_fit}-b and \ref{fig:forces_dft_versus_fit}-c. As expected, at such short distances, the interactions are energetically unfavorable, leading to a low probability of visiting those points in the phase space during a simulation. For example, at \SI{2.5}{\angstrom} from the center of the (8,8) nanotube, when the fit starts to diverge from DFT (Figure \ref{fig:forces_dft_versus_fit}-C), the interaction energy is \SI{2.56}{\kilo\calorie\per\mol}, while the lowest energy at the same adsorption configuration is \SI{-8.02}{\kilo\calorie\per\mol}. Despite describing the force on nitrogen atoms, the fit could not reproduce all the features of the forces on carbon atoms. However, the difference is small and might not lead to distinct results in molecular dynamics simulations due to the thermal movement and the larger contribution of bonded forces acting on such atoms. The oscillation present in the forces from DFT is due to the eggbox effect present in our DFT calculations \cite{Artacho_2008}.

\subsection{Grand Canonical Monte Carlo Simulations}
\begin{figure}[ht]
\includegraphics[width=\columnwidth]{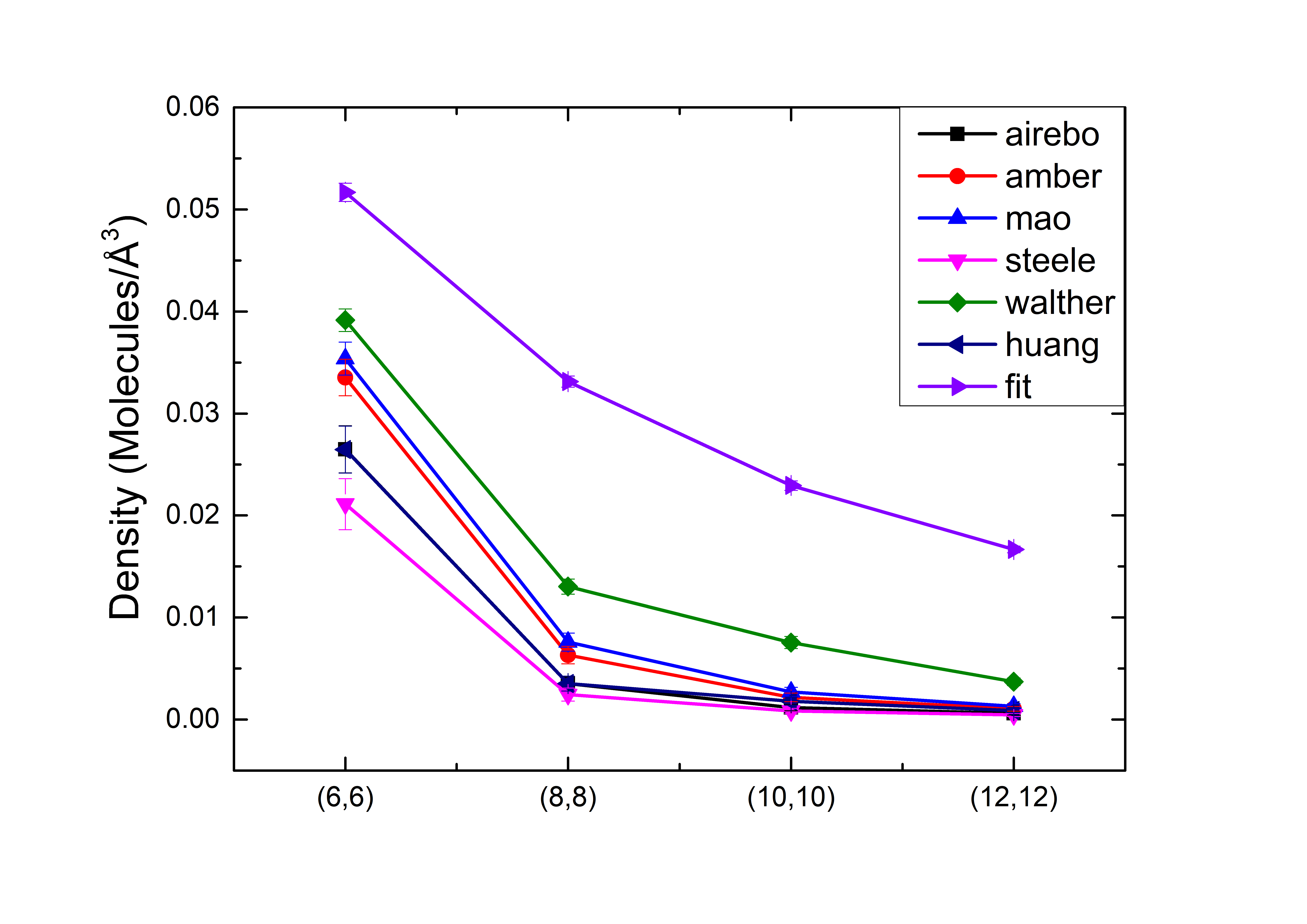}
\caption{\label{fig:n2_load} Comparison of the density of nitrogen molecules inside a (6,6), (8,8), (10,10) and (12,12) carbon nanotubes. The proposed potential predicts a larger number of molecules inside the nanotubes in all cases. }
\end{figure}

Figure \ref{fig:n2_load} compares the density of nitrogen molecules inside different CNTs, using classical force fields with LB and the proposed potential. Each FFs predicts a different density, illustrating the importance of a careful selection of the most appropriate force field for molecular simulations. For all the nanotube diameters, our fit predicts a larger number of nitrogen molecules, reaching an almost \num{40} fold difference for the (12,12) CNT compared to Steele's potential, respectively. The higher number of adsorbed molecules predicted by the fit can be related to the largest depth of the potential well of the carbon-nitrogen interaction and the $\sigma$, which is the distance where the potential is zero, being the smallest. Since our fit agrees with DFT - KBM results, we suggest that the use of classical force fields and Lorentz-Berthelot combination rules underestimate the vdW interaction between a nitrogen molecule and carbon nanotubes, predicting different results than those obtained with a good vdW description.  

\begin{figure}[ht]
\includegraphics[width=\columnwidth]{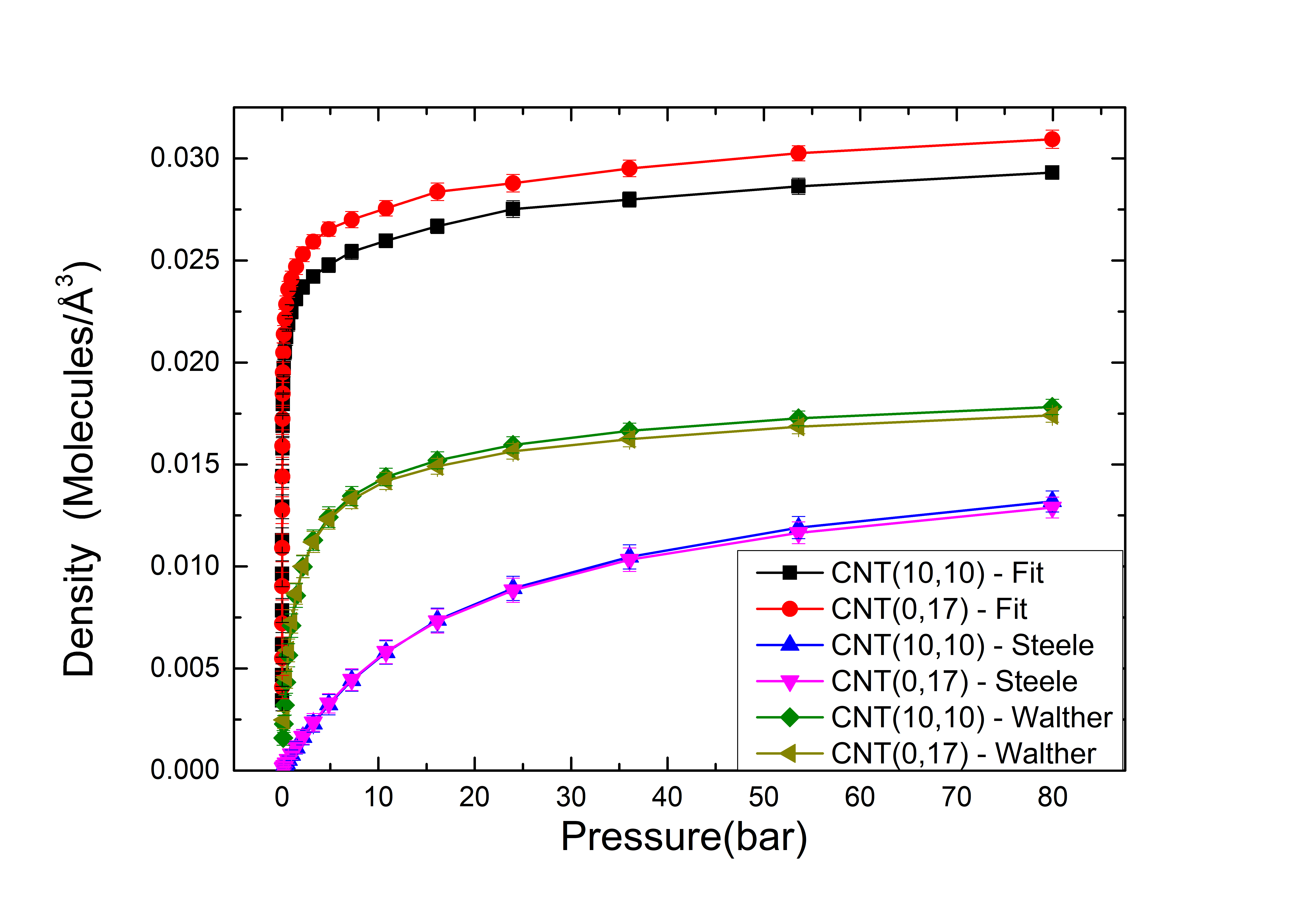}
\caption{\label{fig:isotherms} Adsorption isotherms of \ce{N2} in the (10,10) and (0,17) CNTs, predicted by Steele potential and LB and by the proposed potential. The y-axis is the number of adsorbed molecules inside the simulated CNTs and this number is higher for the fitted potential, due to the larger potential depth.}
\end{figure}

Figure \ref{fig:isotherms} shows the isotherms of CNT (10,10) and (0,17), which have similar radius, obtained using the fitted, Steele's and Walther's potential. There is only small difference between the isotherms of CNT(10,10) and CNT(0,17), indicating that chirality does not play a role in the adsorption properties of carbon nanotubes. The difference is probably due to the slight differences in the volume of the CNTs.

Following the procedure described in Kumar \textit{et al.} \cite{kumar2019}, each potential predicts a different isotherm curve, described by a different isotherm model. As can be seen in Figure~S3, the isotherms predicted by the Steele potential are modeled by a Langmuir isotherm given by equation \ref{eq:languir}, 
\begin{equation}
\label{eq:languir}
q_h=\frac{NKP}{1+KP}
\end{equation}
in which $q_h$ is the concentration of adsorbed molecules, $N$ is the number of adsorption sites, $K$ is related to the binding energy, and $P$ is the pressure. The isotherm predicted from Walther's potential is modeled by a combination of Langmuir equations, as can be seen in the plot of a Scatchard equation on Figure~S4. The proposed potential predicts that the isotherms are modeled by a Freundlich isotherm, as shown in Figure~S5 of the supporting information, which agrees with experimental data in the literature \cite{khalili2013,rahimi2013,erdogan2019}. The Freundlich model for adsorption isotherms is written in equation \ref{eq:freundlich}.
\begin{equation}
\label{eq:freundlich}
q_h=q_mbP^n
\end{equation}
The term $q_m$ is the maximum adsorption capacity, and b is an isotherm constant \cite{kumar2019}. 

The reason for each potential predicts a different isotherm model might be the underestimation of the difference on the adsorption energies of different adsorption sites, since a single Langmuir equation predicts only a adsorption energy, while the Freundlich predicts a distribution. Indeed, the difference in energy between configuration A and C predicted from Steele's potential is \SI{0.002}{\kilo\calorie\per\mol}, while the difference from the fitted potential is \SI{0.19}{\kilo\calorie\per\mol}, both of which below the thermal energy at room temperature ( \SI{0.593}{\kilo\calorie\per\mol}). Therefore, the results suggest that the underestimation of the adsorption energy from the LB combination rules also leads to an underestimation of the energy difference among the different adsorption sites, changing the adsorption properties of a material.

\section{Conclusion}
In this work, we used DFT with vdW to describe the interactions between \ce{N2} with graphene and carbon nanotubes. The comparison of the interaction between nitrogen molecules and carbon nanomaterials obtained through Lorentz-Berthelot combination rules and DFT - KBM calculations show that they underestimate the energies regardless of the potential used for the carbon nanostructures. Because of this underestimation, we fitted a new Lennard-Jones potential. The potential, derived from data from DFT, was aimed to describe \ce{N2} interaction with carbon atoms on graphene and carbon nanotubes. We tested the difference of our proposed potential and Lorentz-Berthelot combination rules on the number of nitrogen molecules inside arm-chair CNTs of different diameters and on the adsorption isotherms. These results suggest that Lorentz-Berthelot rules underestimate the load, making the results obtained using such approximation to be correspondent to those of a lower temperature or lower pressure. Moreover, they also indicate that LB predicts a different curve for the adsorption isotherm due to the underestimation of the difference in adsorption energies, leading to divergent adsorption properties. The parametrizations brought can help in the improvement of the accuracy of molecular simulations in which \ce{N2} molecules are interacting with carbon nanostructures, such as simulations of gas separations.

\section*{Acknowledgments}
CAMJ thanks CAPES for the PhD scholarship. The authors gratefully acknowledge the support of the RCGI – Research Centre for Gas Innovation, hosted by the University of São Paulo (USP) and sponsored by FAPESP – São Paulo Research Foundation (2014/50279-4, 2020/15230-5, and project number 2020/01558-9) and Shell Brasil, and the strategic importance of the support given by ANP (Brazil’s National Oil, Natural Gas, and Biofuels Agency) through the R\&D levy regulation. The authors also acknowledge National Council for Scientific and Technological Development (CNPq) through grant 307064/2019-0 for financial support. HMC thanks to the financial support of FAPESP, grants \#2019/21430-0 and \#2017/02317-2, and the Research Council of Norway through the Centre of Excellence \textit{Hylleraas Centre for Quantum Molecular Sciences} (grant number 262695). The computational time for the calculations was provided by High-Performance Computing facilities at the University of de São Paulo (USP).

\section*{Supporting Information}
The Supporting Information of this work can be found online.

\bibliography{apssamp}

\end{document}